\documentclass[prd,eqsecnum,showpacs,aps,nofootinbib]{revtex4}
\usepackage{latexsym}
\usepackage{graphicx}
\begin{document}
%\draft
\title{Second order gravitational effects on CMB temperature 
anisotropy \\ in $\Lambda$ dominated flat universes}   
\author{Kenji Tomita }
\affiliation{Yukawa Institute for Theoretical Physics, 
Kyoto University, Kyoto 606-8502, Japan}
\author{Kaiki Taro Inoue}
\affiliation{Department of Science and Engineering, Kinki University,
Higashi-Osaka, 577-8502, Japan}  
\date{\today}

\begin{abstract}
We study second order 
gravitational effects of local inhomogeneities on the 
cosmic microwave background radiation in flat universes
with matter and a cosmological constant $\Lambda$.
We find that the general 
relativistic correction to the Newtonian approximation 
is negligible at second order 
provided that the size of the inhomogeneous region is 
sufficiently smaller than the horizon scale.
For a spherically symmetric top-hat type quasi-linear perturbation, the 
first order temperature fluctuation corresponding to the linear integrated
Sachs-Wolfe (ISW) effect is enhanced(suppressed) by the  
second order one for a compensated void(lump). As a function of redshift of
the local inhomogeneity, the second order temperature fluctuations  
due to evolution of the gravitational potential have a peak before 
the matter-$\Lambda$ equality epoch for a fixed comoving size and a density
contrast.  The second order gravitational effects from 
local quasi-linear inhomogeneities at a redshift $z\sim 1$ may significantly 
affect the cosmic microwave background.

\end{abstract}

\maketitle
%\pacs{Valid PACS appear here.
%{\tt$\backslash$\string pacs\{\}} should always be input,
%even if empty.}

%\narrowtext

%ch1 ----------------------------------------------------------
\section{Introduction}
\label{sec:level1}

Generation of temperature anisotropy in the Cosmic Microwave Background (CMB)
due to linear pertubatons of gravitational potentials is called the 
Sachs-Wolfe (SW) effect\cite{sw}. There are two contribution
to the SW effect. The first one called the non-integrated
Sachs-Wolfe effect is produced by fluctuations of gravitational 
potentials at the surface of last scattering. 
The second one called the integrated
Sachs-Wolfe (ISW) effect is generated by 
time-varying gravitational potentials as the CMB photons 
pass through them from the recombination epoch to the
present epoch. 

Recently, much attention has been paid to 
the non-linear version of the ISW effect 
called the Rees-Sciama (RS) effect\cite{rs} on cosmological scales,
since the RS effect of local inhomogeneities at quasi-linear regime
may explain the origin of observed anomalies 
\cite{teg1,oli1,vie1,cru1,eri1,han1} in the 
large-angle CMB anisotriopy \cite{cs,is1}. 
It is argued that the anisotropy for 
compensated asymptotically expanding 
local voids can be larger because the second order effect
enhances the ISW effect \cite{is2}. 
For compensated local lumps, the nature of the second order effect
at the $\Lambda$ dominated epoch remains unknown.

So far, two types of treatment have been used for deriving the RS effect
of local inhomogeneities. One is the 
general relativistic treatment \cite{tom1,tom2,tom3},
in which a non-linear version of the SW effect at second order 
has been studied.  Another one is the simplified
Newtonian treatment \cite{mart0,mart4,tl,tla,
coo1,coo2,coo3} in which influences of nonlinear halo
clustering upon the CMB temperature fluctuations and gravitational 
lensing phenomena have been explored. 

In this paper, we first study the RS effect comparatively
using the general relativistic second order perturbation theory and the
Newtonian approximation and we show the consistency between the two 
approaches in the perturbative regime.
Next, we consider the RS effect of a compensating quasi-linear 
void/lump modelled by a spherically
symmetric top-hat type density perturbation at a redshift $z$   
and we investigate the temporal change
in the linear and second order temperature fluctuations as functions of $z$. 
In \S 2, we derive the solutions of Einstein equations 
for general relativistic perturbations 
at second order in the Poisson gauge (generalized longitudinal
gauge) and consider the limit of $\kappa \ (\sim$ the scale of
inhomogeneities / Hubble radius $) \ll 1$. In \S 3, we study the
perturbations in cosmological Newtonian approximations and their
relation to the general relativistic perturbations up to second order.  
In \S 4, we study the temperature anisotropy owing to first and second
order perturbations based on relativistic perturbation theory
\cite{cmb}.
In \S 5, we investigate the correlations between first order and second order 
temperature fluctuations and their temporal behavior for a spherical
top-hat type density perturbation. 
 \S 6 is dedicated to concluding remarks. In  
Appendices A and B, the main components of Einstein equations and the
derivation of their solutions are shown. In Appendix C, the integrations 
of metric perturbations for spherically symmetric top-hat type perturbations 
along the light paths are shown.

%ch2 ----------------------------------------------------------
\section{General relativistic second order perturbations}
\label{sec:level2}
In what follows, we use the units of $8\pi G = c = 1$, 
the Greek and Latin letters denote $0,1,2,3$ and $1,2,3$,
respectively. Index ``B'' represents the value for the unperturbed background
spacetime. $\delta_{ij} (= \delta^i_j = \delta^{ij})$ are the Kronecker
delta, and subscripts $(n)$ correspond to $n$-th order quantities.

As a function of conformal time $\tilde{\eta}$ and Cartesian coordinates 
$\tilde{x}^i$, the metric of spatially flat Friedmann-Robertson-Walker (FRW) 
universes with first order and second order perturbations 
$\psi^{(n)}, \phi^{(n)},z_i^{(n)},\chi_{ij}^{(n)},n=1,2$ is
described by
\begin{eqnarray}
  \label{eq:m1}
 ds^2 \equiv g_{\mu\nu} d\tilde{x}^\mu d\tilde{x}^\nu &=& \tilde{a}^2 (\tilde{\eta}) \Big\{ 
-(1+ 2 \psi^{(1)} +\psi^{(2)}) d \tilde{\eta}^2 + \Bigl[z_i^{(1)} +{1 \over 2}
z_i^{(2)}\Bigr] d\tilde{\eta} d\tilde{x}^i \cr
&+& \Bigl[(1- 2\phi^{(1)}- \phi^{(2)})\delta_{ij} +
\chi_{ij}^{(1)}+ {1 \over 2}\chi_{ij}^{(2)}\Bigr]d\tilde{x}^i d\tilde{x}^j \Big\} ,
\end{eqnarray}
where $\tilde{a}(\tilde{\eta})$ is the scale factor, 
$\tilde{\eta} \ (=\!\tilde{x}^0)$ is related to the cosmic time 
$t$ by $dt = \tilde{a}(\tilde{\eta}) d\tilde{\eta}$, and $\chi^{(n)}_{ij}$ satisfy
$\chi^{(n)l}_l \equiv \delta ^{lm} 
\chi^{(n)}_{lm} =0$. In what follows, we use the normalized 
scale factor $a=\tilde{a} H_0^{-1}$, 
the comoving coordinate $x^i=\tilde{x}^i/ H_0^{-1}$, 
and the conformal time $\eta=\tilde{\eta}/ H_0^{-1}$ where 
$H_0^{-1}$ is the Hubble radius at present $\eta_0$.
The scale factors at present are defined as $\tilde{a}_0=1$ 
and  $a_0=H_0^{-1}$. 
Note that the scale fatcor $a$ has a dimension of length while
$\eta$ and $r$ do not have dimensions in our notation.

Energy density and $4$-velocity of dust matter are written in terms of
the background quantities and perturbations as  
\begin{equation}
  \label{eq:m2}
\rho = \rho_B(\eta) + \delta^{(1)}\rho +{1 \over
2}\delta^{(2)}\rho, 
\end{equation}
\begin{equation}
  \label{eq:m3}
u^\mu = {1 \over a} \Bigl[\delta^\mu_0 + v^{(1)\mu} + {1 \over
2}v^{(2)\mu}\Bigr], 
\end{equation}
where $v^i$ denotes the 3-velocity of the dust matter.
From the condition $g_{\mu\nu} u^\mu u^\nu = -1$, we obtain the
relations
\begin{equation}
  \label{eq:m4}
v^{(1)0} = - \psi^{(1)},
\end{equation}
\begin{equation}
  \label{eq:m5}
v^{(2)0} = - \psi^{(2)} +3(\psi^{(1)})^2 +
[2z_i^{(1)}+v_i^{(1)}] v^{(1)i}.
\end{equation}

Einstein equations for the unperturbed background with dust matter
and a cosmolological constant $\Lambda$ are

\begin{equation}
  \label{eq:m6}
3(a'/a)^2 = (\rho_B + \rho_\Lambda) a^2,
\end{equation}
\begin{equation}
  \label{eq:m7}
6(a'/a)' = -(\rho_B - 2\rho_\Lambda) a^2
\end{equation}
where $\rho_\Lambda$ is the energy density of the cosmological constant
$\Lambda$ %$\rho_B a^3 =\ $ const%
and a prime denotes derivative with respect to $\eta$. 

To fix the gauge freedom of perturbations, we adopt the Poisson gauge
defined by 
\begin{equation}
  \label{eq:m8}
\delta^{lm} z^{(n)}_{l,m} = 0 \quad {\rm and}\quad \delta^{lm}
\chi^{(n)}_{kl,m} = 0 
\end{equation}
for $n = 1$ and $2$. This gauge is a generalized version of the
longitudinal gauge which is defined by $z_i^{(n)}= 0$ and
$\chi_{ij}^{(n)} = 0$ \cite{eds}, 
and gives us a metric expression convenient for a Newtonian
interpretation, as well as the longitudinal gauge \cite{ks}. 

In what follows, we consider only scalar-type perturabtions at linear
order. Then, we have
\begin{equation}
  \label{eq:m9}
z_l^{(1)} = 0  \quad {\rm and}\quad  \chi^{(1)}_{kl} = 0
\end{equation}
The Ricci tensor, the Einstein tensor and the energy-momentum tensor
for dust matter are shown in Appendix A. Solving the Einstein equation
we obtain the expression of first order scalar-type perturbations in the
growing mode in terms of functions $P(\eta)$ and $F(\bf x)$
as\footnote{In paper \cite{tom1} (referred as Paper\cite{tom1}), the
first order perturbations in the Poisson gauge were derived by transforming
the solution in the comoving synchronous gauge to that in the
Poisson gauge\cite{gauge,eds}. After the publication several misprints
were found in Eqs. (4.6) - (4.8) of
Paper\cite{tom1}, which should be taken into account for 
deriving expressions (\ref{eq:m10}) - (\ref{eq:m13}). 
Perturbations in the decaying mode were derived in Paper \cite{tom1},
 but they are omitted here.}
\begin{equation}
  \label{eq:m10}
\psi^{(1)} = \phi^{(1)} = -{1\over 2}\Bigl(1 - {a' \over a}P'\Bigr) F,
\end{equation}
\begin{equation}
  \label{eq:m11}
z_i^{(1)} = 0, \quad \chi^{(1)}_{ij} = 0,
\end{equation}
\begin{equation}
  \label{eq:m12}
\delta \rho^{(1)}/\rho_B = {1 \over \rho_B a^2} [(a'/a)P' -1] \Delta F +
{3\over 2}(a'/a) P'F,
\end{equation}
\begin{equation}
  \label{eq:m13}
v^{(1)0} = -{1 \over 2}[(a'/a)P' -1] F, \quad v^{(1)i} = {1 \over
2}P'F_{,i}, 
\end{equation}
where $F_{,i}$ is $\partial F/\partial x^i$, \ $\Delta$ is the
Laplacian $\partial^2/\partial x^i\partial x^i$, and   
\begin{eqnarray}
  \label{eq:m14a}
P(\eta) &=& -{2 \over 3\Omega_{m0}}\tilde{a}^{-3/2} [\Omega_{m0}+\Omega_{\Lambda 0} 
\tilde{a}^3]^{1/2}
\int^{\tilde{a}}_0 d \tilde{a} \tilde{a}^{3/2}[\Omega_{m0}+\Omega_{\Lambda 0}
 \tilde{a}^3]^{-1/2} +
{2 \over 3\Omega_{m0}}\tilde{a}, \cr
\eta &=& \int^{\tilde{a}}_0 
d\tilde{a} \tilde{a}^{-1/2}[\Omega_{m0}+\Omega_{\Lambda 0}
\tilde{a}^3]^{-1/2},  
\end{eqnarray}
where $\Omega_{m0}$ and $\Omega_{\Lambda 0}$ are the 
density parameters for matter and the cosmological constant at present. 
$P(\eta)$ is the solution of the growing mode in equation
\begin{equation}
  \label{eq:m14}
P'' + {2a'\over a} P' -1 = 0.
\end{equation}
Note that the potential function $F({\bf x})$ is related to the 
first order matter density contrast $\epsilon_m$ in comoving slices 
(defined in the comoving synchronous gauge) as
\begin{equation}
  \label{eq:m15}
\epsilon_m = {1 \over \rho_B a^2} [(a'/a)P' -1] \Delta F.
\end{equation}

Next we consider relativistic second order perturbations corresponding
to the first order perturbations in the growing mode. 
From Eqs. (\ref{eq:b9}) - (\ref{eq:b12}) in Appendix B, which are
derived by solving the Einstein equations $\mathop{\delta}_2 G_i^j =
\mathop{\delta}_2 T_i^j$, we have
\begin{equation}
  \label{eq:m16}
\phi^{(2)} = \zeta_1  F_{,l}F_{,l} + \zeta_2 \cdot 100\Psi_0 + \zeta_3
F^2 + \zeta_4 \cdot 100\Theta_0,
\end{equation}
\begin{equation}
  \label{eq:m17}
\psi^{(2)} = \xi_1  F_{,l}F_{,l} + \xi_2 \cdot 100\Psi_0 + \xi_3
F^2 + \xi_4 \cdot 100\Theta_0,
\end{equation}
where
\begin{eqnarray}
  \label{eq:m16a}
\zeta_1  &=& {1 \over 4}P \Bigl(1 - {a' \over a}P'\Bigr), \qquad 
\zeta_2  = \Big\{{1 \over 21} {a' \over a}\Bigl(PP' - {1 \over
6}Q'\Bigr) - {1 \over 18}\Bigl[P + {1 \over 2}(P')^2\Bigr] \Big\},\cr
\zeta_3  &=& {1 \over 4}P'\Big\{{a' \over a}+ \Bigl[-{a'' \over
a}+\Bigl({a' \over a}\Bigr)^2 \Bigr]P' \Big\} , \qquad
\zeta_4  = - {1 \over 3} {a' \over a}P'\Bigl(1 - {a' \over a}P'\Bigr), 
\end{eqnarray}
\begin{eqnarray}
  \label{eq:m17a}
\xi_1 &=& \zeta_1, \qquad  
\xi_2 = \zeta_2, \cr
\xi_3 &=& {1 \over 4}\Big\{4 - 7{a' \over a}P' +\Bigl[-{a''
\over a} + 5\Bigl({a' \over a}\Bigr)^2\Bigr](P')^2 \Big\},  \qquad 
\xi_4 = {1 \over 6}\Bigl\{2 - {6a' \over a}P' + \Bigl[-{2a'' \over a} +
8\Bigl({a'\over a}\Bigr)^2 \Bigr](P')^2 \Bigr\},
\end{eqnarray}
\begin{equation}
  \label{eq:m18}
z_i^{(2)} = P' (1 + P'')  C_i,
\end{equation}
and
\begin{equation}
  \label{eq:m19}
\chi_{ij}^{(2)} = \Bigl[P + {1 \over 2}(P')^2\Bigr] 
D_{ij} +{3 \over 7} P^2\Delta D_{ij} + \delta \chi_{ij},
\end{equation}
where $C_i$ satisfies
\begin{equation}
  \label{eq:m20}
\Delta C_i = \Bigl[- {200 \over 9} \Psi_{0,i} + {1
\over 2} (F_{,l}F_{,l})_{,i} - F_{,i} \Delta F\Bigr],
\end{equation}
and $\delta \chi_{ij}$ in Eq.(\ref{eq:m19}) satisfies
\begin{equation}
  \label{eq:m21}
\Bigl({\partial^2 \over \partial\eta^2} +{2a' \over a}{\partial \over
\partial\eta} -\Delta\Bigr) \delta \chi_{ij} = {3 \over 7} P^2
\Delta^2 D_{ij} + {1 \over 7}\Bigl[P - {5 \over 2}(P')^2\Bigr] \Delta
D_{ij}.
\end{equation}
$\Psi_0$ and $\Theta_0$ are defined as
\begin{equation}
  \label{eq:m22}
\Delta \Psi_0 \equiv {9 \over 200} \Bigl[F_{,kl} F_{,kl} - (\Delta
F)^2\Bigr], \quad \Delta \Theta_0 \equiv \Psi_0 - {3 \over 100}
F_{,l} F_{,l}, 
\end{equation}
and $Q(\eta)$ satisfies
\begin{equation}
  \label{eq:m23}
Q'' + {2a'\over a} Q' = - \Bigl[P - {5\over 2}(P')^2 \Bigr].
\end{equation}
The above second order solutions are
consistent with those shown in 
Eqs.(4.12) - (4.15) in Paper\cite{tom1}, which was derived using a 
transformation from the comoving synchronous gauge to the
Poisson gauge. Note that $\Delta D_{ij}$ and $\Delta^2
D_{ij}$ correspond to $\tilde{G}_i^j$ and $G_i^j$ in Eq.(2.25) of
Paper\cite{tom1}, as $\Delta D_{ij} = - \tilde{G}_i^j$ and $\Delta^2
D_{ij} = -G_i^j$.  Eq.(2.17) of Paper\cite{tom1} with misprints must
be replaced by the above correct equation (\ref{eq:m23}).  

In the above solutions, the ratios of terms including $\Theta_0$ to
terms including $F_{,l}F_{,l}$ and $\Psi_0$ are of the order of
$\kappa^2$, where $\kappa \equiv |{\bf
 x}|/\eta$, and $|{\bf x}|$ and $\eta$ are the characteristic spatial
scale of local perturbation and the horizon size. Therefore, the 
terms including $\Theta_0$ are negligible for local 
perturbations that are sufficiently smaller than the horizon size. 
Thus for $\kappa \ll 1$, we have    
\begin{equation}
  \label{eq:m24}
\phi^{(2)} = \psi^{(2)} = \zeta_1 \ F_{,l}F_{,l} + \zeta_2
\ 100\Psi_0,
\end{equation}
\begin{equation}
  \label{eq:m27}
z_i^{(2)} = P' (1 + P'')  C_i,
\end{equation}
and
\begin{equation}
  \label{eq:m28}
\chi_{ij}^{(2)} = {3 \over 7} P^2\Delta D_{ij}.
\end{equation}
%

%ch3 ----------------------------------------------------------
\section{Cosmological Newtonian approximation}
\label{sec:level3}

In the cosmological Newtonian approximation, we assume that
$\epsilon \equiv a |{\bf v}| /c \ll 1$ and $\kappa \equiv |{\bf x}|
/\eta \ll 1$ but $\chi \equiv (\rho/\rho_B -1)^{1/2}$ is arbitrary, and
consider only $\psi$ and $\phi (=\psi)$ as the
metric perturbations\cite{newt}. Then, from the difference 
between the perturbed equation $R^0_0 = T^0_0 - {1 \over 2} T^\mu_\mu$ 
and the background counterpart, we obtain 
\begin{equation}
  \label{eq:r1}
\Delta \psi = {1 \over 2} a^2 (\rho - \rho_B).
\end{equation}
From the conservation equation $T^{\mu\nu}_{;\nu} = 0$ and the
energy-momentum tensor (for the perfect fluid with pressure $p$ )
\ $T^{\mu\nu} = 
(\rho + p) u^\mu u^\nu + p g^{\mu\nu}$, we obtain the equation of
continuity 
\begin{equation}
  \label{eq:r2}
\rho' + {3a' \over a} \rho + (\rho v^i)_{,i} = 0
\end{equation}
and the equation of motion
\begin{equation}
  \label{eq:r3}
{v^i}' + v^i_{,j} v^j +{a' \over a}v^i -\psi_{,i} +
p_{,i}/\rho  = 0,
\end{equation}
where $v^i = a u^i$. By solving these equations (\ref{eq:r1})
-(\ref{eq:r3}), we have $\rho,  v^i$ and 
$\psi$, which can determine the lowest-order RS effect in
the form of spatial integration of $\psi_{,i}$ along a
light path. 

In what follows,  we only consider the perturbative case with $\epsilon \ll
1$, $\kappa \ll 1$, and  $\chi \ll 1$. First, expressing the
perturbations as
\begin{equation}
  \label{eq:r4}
\psi=\psi^{(1)} + {1\over 2}\psi^{(2)}, \qquad \rho=\rho_B +
\delta \rho^{(1)} +{1\over 2}\delta \rho^{(2)}, \qquad v^i = {v^{(1)}}^i
+{1\over 2} {v^{(2)}}^i,
\end{equation}
the first order equations for these perturbations in the
pressureless case are given by
\begin{eqnarray}
  \label{eq:r5}
\Delta \psi^{(1)} &-& {1\over 2}a^2 \delta \rho^{(1)} = 0, \cr
 \delta {\rho^{(1)}}' &+& {3 a' \over a}\delta \rho^{(1)} +\rho_B
{{v^{(1)}}^i}_{,i} =0, \cr
{{v^{(1)}}^i}' &+& {a' \over a}{v^{(1)}}^i + \psi^{(1)}_{,i} =0 ,
\end{eqnarray}
and the first order solutions are 
\begin{equation}
  \label{eq:r6}
\delta \rho^{(1)}/\rho_B = {1 \over \rho_B a^2} [(a'/a)P' -1] \Delta F,
\end{equation}
\begin{equation}
  \label{eq:r7}
v^{(1)i} = {1 \over 2}P'F_{,i}, 
\end{equation}
and
\begin{equation}
  \label{eq:r8}
\psi^{(1)} = -{1\over 2}\Bigl(1 - {a' \over a}P'\Bigr) F,
\end{equation}
where the above $\delta \rho^{(1)}, v^{(1)i}$ and $\psi^{(1)}$ are equal to
Eqs.(\ref{eq:m12}), (\ref{eq:m13}), and (\ref{eq:m10}) in the limit of
$\kappa \ll 1$, respectively.
 
Next the corresponding Newtonian second order equations are
\begin{eqnarray}
  \label{eq:r9}
\Delta \psi^{(2)} &-& {1\over 2}a^2 \delta \rho^{(2)} = 0, \cr
\delta {\rho^{(2)}}' &+&{3 a' \over a}\delta \rho^{(2)} + 2[\delta
\rho^{(1)} {v^{(1)}}^i]_{,i} + \rho_B {v^{(2)}}^i_{,i} = 0, \cr
{{v^{(2)}}^i}' &+& 2\ {v^{(1)}}^i_{,j} {v^{(1)}}^j + {a' \over
a}{v^{(2)}}^i +\psi^{(2)}_{,i} = 0.
\end{eqnarray}
By substituting the above first order solutions 
Eqs. (\ref{eq:r6}), (\ref{eq:r7}), and (\ref{eq:r8}) to Eq.(\ref{eq:r9})
and eliminating $\psi^{(2)}$ and ${v^{(2)}}^i$, we obtain
\begin{equation}
  \label{eq:r10}
(\delta \rho^{(2)}/\rho_B)'' +{a' \over a}(\delta \rho^{(2)}/\rho_B)'
- {1\over 2}a^2 \delta \rho^{(2)} = {1\over 4}(P')^2 \Delta
(F_{,l} F_{,l}) - {1 \over \rho_Ba^4}[a^2 P' ({a' \over a}P' -1)]'
(\Delta F F_{,i})_{,i}.
\end{equation}
Assuming that $\psi^{(2)}$ is given by Eq.(\ref{eq:m24}) and
using the first line of Eq.(\ref{eq:r9}), we can express $\delta \rho^{(2)}$ as
\begin{equation}
  \label{eq:r11}
\delta \rho^{(2)}/\rho_B = {2\over \rho_Ba^2} \{\zeta_1 \Delta
(F_{,l} F_{,l}) + {9\over 2}\zeta_2 [F_{,ij}F_{,ij} -(\Delta F)^2] \}.
\end{equation}
By substituting this $\delta \rho^{(2)}/\rho_B$ to Eq.(\ref{eq:r10}), we
find that it is a solution of Eq.(\ref{eq:r10}), so that $\psi^{(2)}$
given by Eq.(\ref{eq:m24}) is also a solution of Eq.(\ref{eq:r9}).
Thus we proved that in the cosmological Newtonian limit 
with $\epsilon \ll1$, $\kappa \ll 1$, and  $\chi \ll 1$, 
the general relativistic
second order solution is consistent with the Newtonian one.

%ch4 ----------------------------------------------------------
\section{Temperature anisotropy}
\label{sec:level4}

In this section we consider the observed temperature of the CMB
radiation which was emitted at the recombination epoch and received at
the present epoch. The relation between the 
emitted and received temperatures $T_e$ 
$T_o$ is
\begin{equation}
  \label{eq:c1}
T_o = (\omega_o/\omega_e) T_e, 
\end{equation}
where $\omega = - g_{\mu\nu} u^\mu k^\nu$, $u^\mu$ is the
observer's and emitter's velocities, and $k^\mu \ (= dx^\mu/d\lambda)$ is
the wave vector of photons with affine parameter $\lambda$, which
satisfies the null geodesic equation in the perturbed
FRW universe. Solving this equation, the first order and second order
perturbations of observed temperature $\Delta T^{(1)} $ and
$\Delta T^{(2)}$ were derived in the gauge-invariant manner by
Mollerach and Materrese\cite{cmb}. When the background null geodesic 
rays are expressed 
as $x^{(0)\mu} = (\lambda, (\lambda_o - \lambda) e^i)$ and $k^{(0)\mu}
= (1, -e^i)$, the first order temperature fluctuation is 
\begin{equation}
  \label{eq:c2}
\Delta T^{(1)}/T = \psi^{(1)}_e - \psi^{(1)}_o + [v^{(1)i}_o
-v^{(1)i}_e] e_i + \tau + I_{1e}, 
\end{equation}
where 
\begin{eqnarray}
  \label{eq:c3}
I_{1e} &\equiv& - \int^{\lambda_e}_{\lambda_o} d\lambda {A^{(1)}}',\cr
A^{(1)} &\equiv& \psi^{(1)}+\phi^{(1)}+ z_i^{(1)} e^i - {1\over 2}
\chi_{ij}^{(1)} e^i e^j,
\end{eqnarray}
$e^i$ is the unit (three-dimensional) directional vector,
the subscripts $e$ and $o$ denote the epochs of emission and
observation, $\tau$ is the temperature fluctuation at the emission
epoch, and it is assumed that $x^{(1)\mu} (\lambda_o) = 0$ and
$k^{(1)i} (\lambda_o) =0$. The integral term $I_{1e}$ represents
the contribution due to the ISW effect.  

Similarly, second order temperature fluctuation is given by
\begin{eqnarray}
  \label{eq:c4}
\Delta T^{(2)}/T &=&  I_{2e}+ [I_{1e}]^2  +(\Delta T^{(2)}/T)_{oe},\cr
I_{2e}&=&- {1\over 2}\int^{\lambda_e}_{\lambda_o}
d\lambda {A^{(2)}}'
\end{eqnarray}
where $(\Delta T^{(2)} T/T)_{oe}$ is the sum of terms
consisting of second order and products of first order quantities at
observer's and emitter's positions, and
\begin{equation}
  \label{eq:c5}
A^{(2)} \equiv \psi^{(2)}+\phi^{(2)}+ z_i^{(2)} e^i - {1\over 2}
\chi_{ij}^{(2)} e^i e^j.
\end{equation}
For quasi-linear perturbations with $\epsilon_m={\cal{O}}(0.1)$,
the contribution due to the RS effect can be written as  
$(\Delta T/T)_{{\textrm{\small{RS}}}}=I_{1e}+I_{2e}$. 
\begin{figure}[t]
\caption{\label{fig:dzeta} Time dependent terms $\zeta'_1$ and
 $-\zeta'_2$ in second order temperature fluctuations as a function of
 redshift $z$. $\Omega_{m0}$ denotes the present value for the matter
density parameter.  }
\includegraphics[width=8cm]{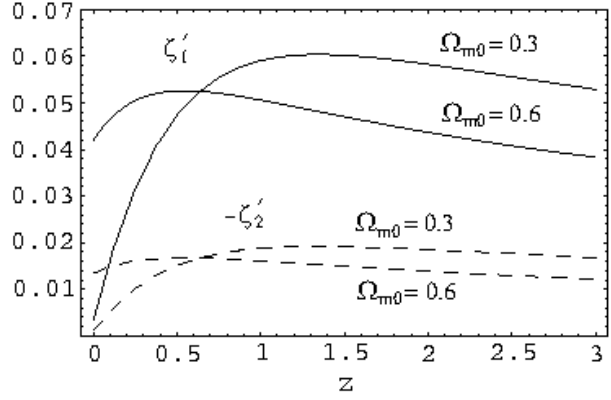}
\end{figure}
For scalar-type linear or quasi-linear perturbations, the terms 
$z_i^{(2)}$ and $\chi_{ij}^{(2)}$ in Eq.(\ref{eq:c5}) 
can be neglected. Therefore, the second order contribution 
$I_{2e}$ to the RS effect can be evaluated using only 
perturbations ${\phi^{(2)}}'$ and ${\psi^{(2)}}'$ which 
are obtained from Eq. (\ref{eq:m24}) with the following
conformal time derivatives
\begin{equation}
  \label{eq:c10}
{\zeta_1}' = {1\over 4} \Big\{P' - {a'\over a}[(P')^2 +P]
+\Bigl[2\Bigl({a'\over a}\Bigr)^2 
-\Bigl({a'\over a}\Bigr)'\Bigr] PP' \Big\}, 
\end{equation}
\begin{equation}
  \label{eq:c11}
{\zeta_2}' = {1\over 18} {a'\over a}P - {1\over 9}P'+ {1\over 21}
\Bigl[\Bigl({a'\over a}\Bigr)' -2\Bigl({a'\over a}\Bigr)^2\Bigr]
\Bigl(PP'- {1\over 6}Q'\Bigr) + {5 \over 
36}{a'\over a} (P')^2, 
\end{equation}
where $z_i^{(2)}$ and $\chi_{ij}^{(2)}$ are neglected.
For perturbations sufficiently smaller than the Hubble scale, 
i.e., $\kappa \equiv|{\bf x}|/\eta \ll 1$, 
the dominant terms in ${\phi^{(2)}}'$ and ${\psi^{(2)}}'$ 
are those multiplied by 
$F_{,l}F_{,l}$ and $\Psi_0$. 

For the Einstein-de Sitter (EdS) model ($\Lambda = 0$), we have $a \propto
\eta^2, P=\eta^2/10$ and $Q = 0$, so that 
\begin{equation}
  \label{eq:c14}
\zeta_1 = {3\over 200} \eta^2, \quad \zeta_2 = -{1\over 210} \eta^2, \
{\rm and} \ \zeta'_1/\zeta'_2 = -{63 \over 20}.
\end{equation}
For a $\Lambda$-dominated model with $(\Omega_{m0}, \Omega_{\Lambda0} 
= (0.27,0.73)$, numerical calculations of $\zeta_1$ and $\zeta_2$ show that 
$\zeta'_1/\zeta'_2$ have the values $-2.68, -2.99, -3.08, -3.12,
-3.14, -3.15$ for the redshifts $z = 0.05, 0.1, 0.2, 0.5, 1.0, 2.0,$
respectively. For $z \gg 1$,  $\zeta'_1/\zeta'_2$ is nearly equal to
$-63/20$ in the EdS model. As shown in Fig. \ref{fig:dzeta}, 
$\zeta'_1$ and $\zeta'_2$ have a peak well before the matter-$\Lambda$
equality epoch $z_{m\Lambda}=((1-\Omega_{m0})/\Omega_{m0})^{1/3}-1$,
implying that the second order contribution to the RS effect in the 
quasi-linear regime is not so important at an accelerating epoch.  
  
%-----------------------------------------------------
\begin{figure}[t]
\caption{\label{fig:rs1} A light path in the spherical model.}
\includegraphics[width=15cm]{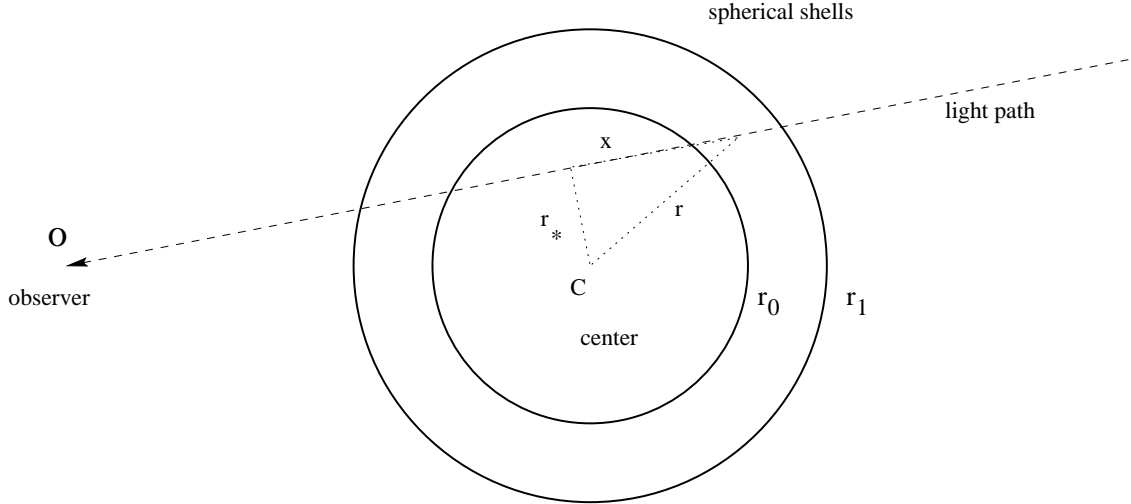}
\end{figure}
%------------------------------------------------------

%-----------------------------------------------------
\begin{figure}[t]
\caption{\label{fig:rs2} The matter density 
contrast for a top-hat type spherical void.}
\includegraphics[width=8cm]{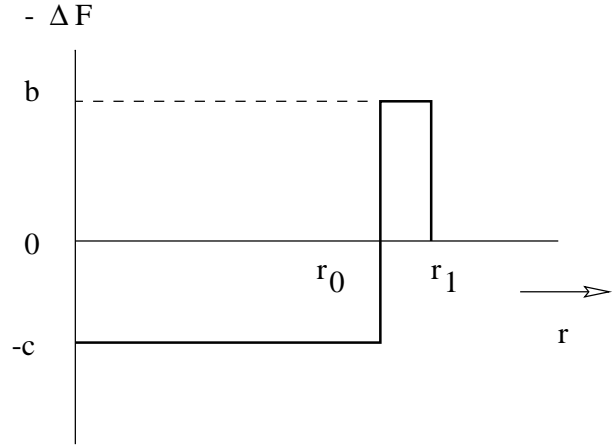}
\end{figure}
%------------------------------------------------------

%ch5 ----------------------------------------------------------
\section{Behaviors of first order and second order temperature
fluctuations in a spherical top-hat model} 

In order to investigate the nature of the temperature fluctuations
exerted by local inhomogeneities, we consider a simple 
toy model with a spherically symmetric density perturbation. 
In what follows, we assume 
that we are outside the local density perturbation (Fig.\ref{fig:rs1}).
In this model, the potential function $F(\textbf{x})$ can be written as a
function of the comoving distance $r \equiv [{\bf  
x}^2]^{1/2}$ from the center of the perturbation as $F = F(r)$. 
Then, we have $C_i = 0, D_{ij} = 0$ and 
$z_i^{(2)} = \chi_{ij}^{(2)} = 0$.  For the 
top-hat type matter density perturbations, the 
functional form for $F(r)$ is given in terms of 
constant parameters $b$ and $c$ as
\begin{equation}
  \label{eq:c15}
\Delta F = {1 \over r^2}{d \over dr} (r^2 F_{,r}) = c,\quad -b, \quad 0 
\end{equation}
for $0 \le r\le r_0,\ r_0<r\le r_1,\ r_1<r$, 
respectively(Fig. \ref{fig:rs2}).
This model represents a void 
if $c > 0$ and $b > 0$, or a lump if $c < 0$ and $b < 0$. 
Moreover, if $c {r_0}^3 = b({r_1}^3 -{r_0}^3)$, 
the mass is totally compensating. 
$-\Delta F$ represents a value that is proportional to 
the matter density contrast in a comoving gauge. 
Note that $-F(\textbf{x})$  describes the $\textbf{x}$ dependence of the 
gravitational potential $\psi$. 
Here, $r_0$ and $r_1$ are the inner 
radius and the outer radius, respectively.  
$r_1-r_0$ corresponds to the width of the wall.

Integrating Eq.(\ref{eq:c15}), we obtain    
\begin{equation}
  \label{eq:c16}
F_{,r} = {1 \over r^2} \int^r_0 dr r^2 \Delta F.
\end{equation}
Then, under the condition that $F_{,r}$ is regular at the center $r = 0$,
we have
\begin{equation}
  \label{eq:c17}
F_{,r} = {1 \over 3}cr, \quad {1 \over 3}(c+b)r_0^3/r^2 -{1 \over
3}br, \quad {1 \over 3r^2}[c r_0^3 - b(r_1^3 - r_0^3)]
\end{equation}
for $0 \le r\le r_0, r_0<r\le r_1, r_1<r$, respectively.
From Eq.(\ref{eq:m22}), on the other hand, we have
\begin{equation}
  \label{eq:c18}
-{100\over 9} (r^2 \Psi_{0,r})_{,r} = [r(F_{,r})^2]_{,r},
\end{equation}
so that 
\begin{equation}
  \label{eq:c19}
-{100\over 9}[\Psi_0(r)) -\Psi_0(0)] = \int^r_0 {(F_{,r})^2 \over r} dr
\equiv I.
\end{equation}
Under the condition that $\Psi_0$ is regular at the center, 
$\Psi_0(0)$ is determined from the boundary 
condition that the perturbation is local, i.e., $\Psi_0 \rightarrow 0$ as $r
\rightarrow \infty$. The explicit forms of $I$ and $\Psi_0(0)$ are shown
in Appendix C. 

\begin{figure}[t]
\caption{\label{fig:F} The potential function $-F(r)$ 
for a top-hat type compensated lump with $c=-0.1$ and $r_1=0.1$.}
\includegraphics[width=9cm]{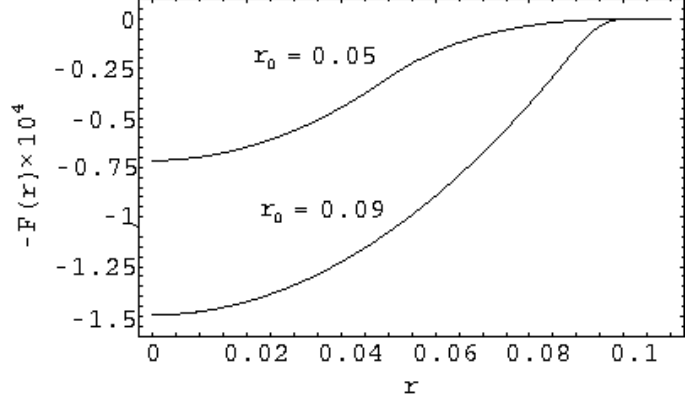}
\end{figure}

Integrating Eq.(\ref{eq:c17}), we obtain 
\begin{eqnarray}
%  \label{eq:5.6}
F &=&-{1\over 6} c \ (r^2-r_0^2) + F_0 \qquad {\rm for} \qquad 0 \le r <
r_0 \cr
&=& -{1\over 3}(c+b)\ r_0^3 \Bigl({1\over r}-{1\over r_0}\Bigr) -{1\over
6}b\ (r^2-r_0^2) + F_0 \qquad {\rm for} \qquad r_0 \le r <r_1 \cr
&=& -{1 \over 3r}[c r_0^3 - b(r_1^3 - r_0^3)] \qquad {\rm for}
\qquad r_1 \le r,
\end{eqnarray}
where
\begin{equation}
%  \label{eq:5.7}
F_0 = {1\over 3}(c +b)\ r_0^3 \Bigl({1\over r_1}-{1\over r_0}\Bigr)
+ {1\over 6}b \ (r_1^2 - r_0^2) -{1 \over 3r_1}[c r_0^3 - b(r_1^3 -r_0^3)].
\end{equation}
As shown in Fig. \ref{fig:F}, the thinner wall (i.e.,$(r_1-r_0)/r_1 \ll 1$ ) 
gives deeper potential for a lump provided that the matter 
density at the center is fixed (i.e., $c$ is constant). 

%------------------------------------------------------

Using Eqs.(\ref{eq:m10}), (\ref{eq:c2}), and (\ref{eq:c3}), the first order
temperature fluctuation due to the local perturbation is given by the
integral term in Eq. (\ref{eq:c2}), which represents the ISW effect,
%F
\begin{equation}
  \label{eq:fs1}
(\Delta T^{(1)}/T)_{loc} = I_{1e} = - 2
\int^{\lambda_e}_{\lambda_o} d\lambda {\phi^{(1)}}',   
\end{equation}
where
\begin{equation}
  \label{eq:fs2}
{\phi^{(1)}}' = {1 \over 2} \Bigl\{{a'\over a} + \Bigl[{a''\over a}
-3\Bigl({a'\over a}\Bigr)^2\Bigr]P' \Bigr\} F.
\end{equation}

\begin{figure}[t]
\caption{\label{fig:void} The first and second order 
temperature fluctuations as functions of the matter density
parameter at present $\Omega_{m0}$ for photons passing 
through the center of a compensated spherical void at $z \sim 0$. 
The matter density contrast at the center is 
$(\epsilon_{m})_c = -0.3$ and $r_0 = 0.09H_0^{-1}, \ r_1 = 0.1
 H_0^{-1}$, where $H_0$ is the Hubble constant.
 The dashed and dashed-dotted curves denote $ (\Delta T^{(1)}/T)_{loc}$
 and $(\Delta T^{(2)}/T)_{loc}$, respectively. 
The solid curve represents the total temperature fluctuation 
$(\Delta T^{(1)}/T)_{loc}+(\Delta T^{(2)}/T)_{loc}$.}
\includegraphics[width=10cm]{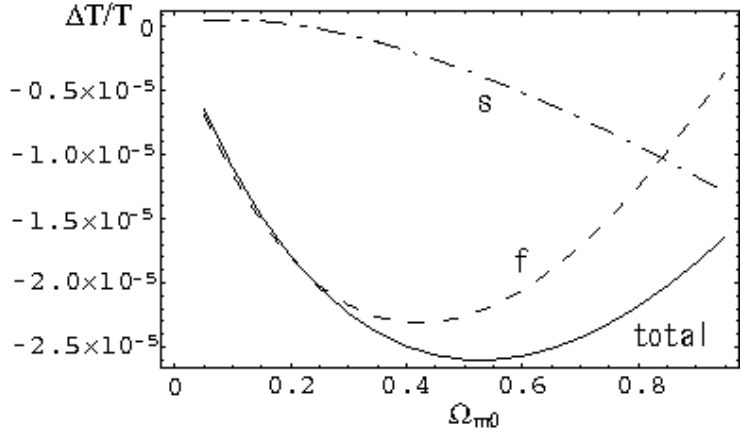}
\end{figure}

\begin{figure}[t]
\caption{\label{fig:lump} For a compensated spherical lump.
The parameters are as the same as in Fig. \ref{fig:void}. 
The matter density contrast is $(\epsilon_m)_c=0.3$.   }
\includegraphics[width=10cm]{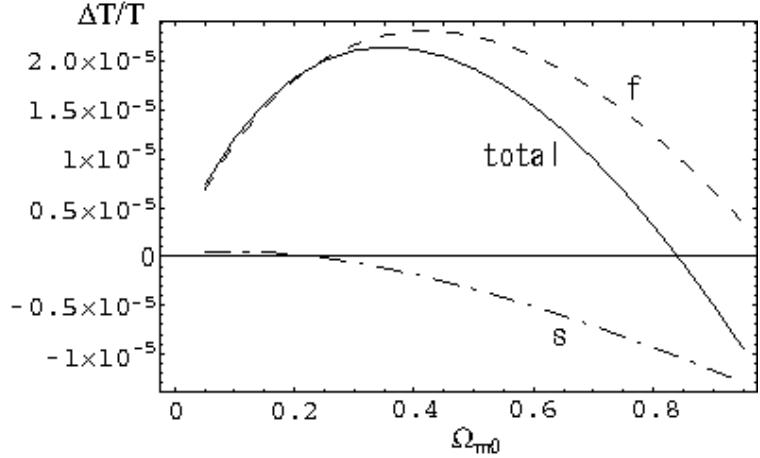}
\end{figure}

Now we consider a light path passing through the center of a spherical
compensating void/lump and assume that its outer radius $r_1$
is sufficiently small compared with the Hubble radius at the time 
the photons pass through the center. Then the integration
of $F$ derived in Appendix C reduces to
\begin{equation}
  \label{eq:fs5}
\int^\infty_0 F dr = -{1\over 9} c (r_0)^3 \ (1+ y) \ln (1+
1/y) 
\end{equation}
%F
where $y = b/c$ and $r_1/r_0 = (1 + 1/y)^{1/3}$.
From the above equations, we can express the first 
order temperature fluctuation as
\begin{equation}
  \label{eq:fs6}
(\Delta T^{(1)}/T)_{loc} \approx {2\over 9}\ c\ (r_1)^3\ w_1(y) \Bigl [
{a'\over a}+\Bigl({a'' \over a} - 3\Bigl({a'\over a}\Bigr)^2\Bigr) P'
\Bigr]\Bigl|_{\eta=\eta_c},   
\end{equation}
where $w_1(y)=-y\ln (1+1/y)$ which is a negative definite function for
$y>0$ and $\eta_0-\eta_c$ represents the comoving distance to the void/lump.
Using Eqs.(\ref{eq:m15}), (\ref{eq:m6}), and (\ref{eq:m7}),
$c$ can be written in terms of the 
matter density contrast $(\epsilon_m)_c$ in the comoving gauge 
at the center of the inhomogeneity as
\begin{equation}
  \label{eq:fs9}
c= (\epsilon_m)_c\ 
2\Bigl[2\Bigl({a'\over a}\Bigr)^2 -{a'' \over a}\Bigr] \Bigl[
{a'\over a}P' -1 \Bigr]^{-1}.
\end{equation}
From Eq.(\ref{eq:fs6}) and Eq.(\ref{eq:fs9}), we find that 
the first order 
temperature fluctuation due to a compensated
void or lump is approximately written as 
${\cal{O}}[(\Delta T^{(1)}/T)_{loc}] \sim (\epsilon_m)_c r_1^3$.   
As shown in Figs. \ref{fig:void} and \ref{fig:lump}  a compensated
spherical void redshifts the photons whereas a compensated spherical lump
blueshifts the photons at first order irrespective of the width of the wall.

In a similar manner, we can express the 
second order temperature fluctuation when
the photons pass through the center of a compensating void/lump
$(= - {1\over 2}
\int^{\lambda_e}_{\lambda_o} d\lambda {A^{(2)}}'$) 
as 
\begin{equation}
  \label{eq:fs7}
(\Delta T^{(2)}/T)_{loc} \approx {4\over 27}\ c^2\ (r_1)^3\ w_2(y)
(\zeta_1 + 9\ \zeta_2)'|_{\eta=\eta_c}, 
\end{equation}
\begin{equation}
  \label{eq:fs8}
w_2(y) \equiv y[1- y \ln(1+1/y)],
\end{equation}
where $c$ is given by Eq.(\ref{eq:fs9}). Note that $w_2(y)$ is a
positive definite function for $y>0$.
The order of the second order temperature fluctuation is 
${\cal{O}}[(\Delta T^{(2)}/T)_{loc}]\!\sim\! -(\epsilon_m)_c^2 r_1^3$. 
As shown in Figs. \ref{fig:void} and \ref{fig:lump}, 
either type (void or lump) of density perturbation 
redshifts the photons irrespective of the width of the wall.
This behavior suggests that the second order gravitational effect 
leads to a flow of matter from the wall to inside the wall
in either case. Then the gravitational potential becomes smaller
and photons passing through the center of the 
void(lump) get further redshifts.

%--------------------------------------------------------
%------------------------------------------------------
\begin{figure}[t]
\caption{\label{fig:rees7} The ratio between the second and first order 
temperature fluctuations $U$ for a
light path passing through the center as a function of
the redshift $z$ of the void/lump. 
The solid, dashed, and dashed-dotted curves correspond to
flat models with  $\Omega_{m0}=0.9, 0.6$, and 0.3, respectively. 
The width of the wall is chosen to be 20\% of the outer radius, i.e.,
 $(r_1-r_0)/r_1=0.2$. It turns out that the 
dependence on the ratio between the width and the outer 
radius is not prominent. }

\includegraphics[width=10.5cm]{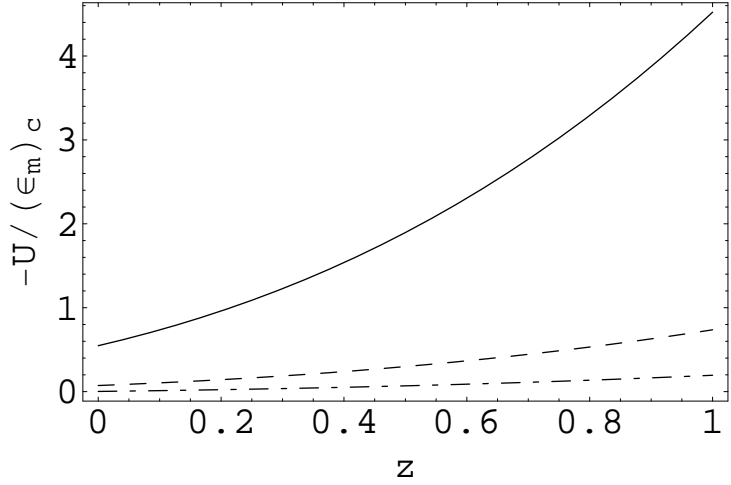}
\end{figure}

To see this quantitatively, 
we calculate the ratio $U$ of the second order temperature fluctuation
to the first order one for a spherical void(lump) at redshift $z$,
\begin{equation}
  \label{eq:fs10}
U \equiv (\Delta T^{(2)}/T)_{loc}(z)/ (\Delta T^{(1)}/T)_{loc}(z), 
\end{equation}
which is proportional to $c$. Plugging Eq.(\ref{eq:fs9}) into 
Eq.(\ref{eq:fs10}), we have
\begin{equation}
  \label{eq:fs11}
U = {w_2(y) \over w_1(y)} {\ \Bigl[2\Bigl({a'\over
a}\Bigr)^2 -{a'' \over a}\Bigr] \ (\zeta_1 + 9\zeta_2)' \over 6
\Bigl({a'\over a}P' -1\Bigr) \ \Bigl\{{a'\over a}+\Bigl[{a''\over a} - 
3\Bigl({a'\over a}\Bigr)^2\Bigr] P' \Bigr\}}
(\epsilon_m)_c. 
\end{equation}
As shown in Fig. \ref{fig:rees7}, the absolute value 
$|U|$ for a fixed value $(\epsilon_m)_c$ monotonically increases as
the redshift $z$ of a void/lump increases. 
This behavior is naturally expected, since all the 
models approach to the EdS model for which $U$ is infinite 
(because of ${\phi^{(1)}}' = 0$) in the limit $z\rightarrow \infty$. 
It turns out that the ratio $U$ is positive  
(negative) for $(\epsilon_m)_c < 0 (> 0)$, respectively. In other words,
we have positive (negative) correlation for a void (lump) between the
first and the second order fluctuations. 
Thus, we conclude that \textit{the 
first order temperature fluctuation corresponding to the linear integrated
Sachs-Wolfe (ISW) effect is  enhanced(suppressed) by the  
second order effect for a void(lump)}. 

\begin{figure}[t]
\caption{\label{fig:rees4} The first order and second
order temperature fluctuations as a function of the 
redshift $z$ of a void/lump with wall width $(r_1-r_0)/r_1=0.2$ 
, a density contrast parameter $c$, and density parameters 
$(\Omega_{m0},\Omega_{\Lambda0})=(0.27,0.73)$. 
The solid and dashed curves denote $0.05\ (\Delta
 T^{(2)}/T)_{loc}(z)
/(\Delta T^{(2)}/T)_{loc}(0)$ and $(\Delta T^{(1)}/T)_{loc}(z)
/(\Delta T^{(1)}/T)_{loc}(0)$, respectively. }
\includegraphics[width=10cm]{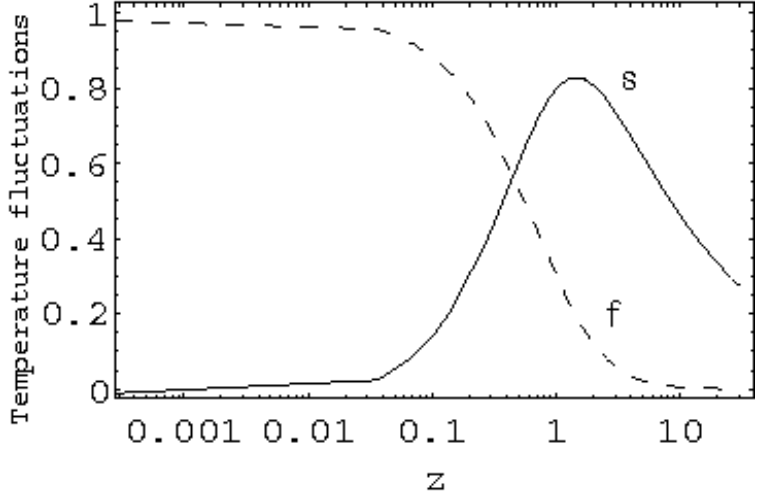}
\end{figure}

For $(r_1-r_0)/r_1=0.2$, or equivalently, $y=b/c=1.049$, the behaviors of
$(\Delta T^{(1)}/T)_{loc}(z)/(\Delta T^{(1)}/T)_{loc}(0)$
and $0.05\ (\Delta T^{(2)}/T)_{loc}(z)/(\Delta T^{(2)}/T)_{loc}(0)$ 
are shown in Fig. \ref{fig:rees4}, when $c$ 
is fixed. In contrast to the first order fluctuations which decrease as
$z$ increases, it is found that the second order fluctuations have a
peak at a certain epoch $z_p$. For a model with  $(\Omega_{m0},
\Omega_{\Lambda 0}) = (0.27, 0.73)$, we have $z_p \sim 1$. 

%--------------------------------------------------------
%------------------------------------------------------

\section{Concluding remarks}

In this paper we have confirmed that
in the nonzero-$\Lambda$ cosmological
model, the second order solutions (in the
Poisson gauge) of scalar-type perturbations 
in general relativistic theory
coincide with the solutions in the cosmological Newtonian theory 
in the 'Newtonian' limit, provided that the size of the perturbation is 
sufficiently small compared with the horizon scale.
Thus, the second order temperature fluctuation due
to a local perturbation sufficiently smaller than the 
horizon scale can be calculated using 
the Newtonian approximation as long as the solutions satisfying
the Poisson equation and equations of continuity and motion are used
in the Newtonian approximation. 

In order to clarify the behavior of second order temperature
fluctuations, we considered a simple spherical top-hat type 
local density perturbation (void/lump) in which the pressureless 
matter is totally compensating and whose
size is smaller than the horizon size. We have found that the 
first order temperature fluctuation corresponding to the linear integrated
Sachs-Wolfe (ISW) effect is enhanced(suppressed) by the  
second order effect for a compensated void(lump). 
As a function of redshift $z$ of the 
local perturbation, the amplitude of the second order 
temperature fluctuation due to a void(lump) 
has a peak at an epoch $z\sim 1$ 
for a fixed density contrast and the comoving size, 
whereas the amplitude of the 
first order fluctuation increases monotonically. 
This means that the second order temperature
fluctuations may play an important role 
especially for quasi-linear objects such as voids with radius $r= 100-200
h^{-1} \textrm{Mpc}$ at $z\sim 1$, which may lead to some observable imprints at 
an angle $\sim 5^{\circ }$ in the sky 
(\cite{is1}, \cite{is2}, \cite{rudnick2007}).

% appA
%----------------------------------------------
\appendix
\section{The main components of first order and second order Einstein
equations} 

The first order and second order components of Ricci and Einstein
tensors are expressed using the symbols $\displaystyle \mathop{\delta}_1$ and
$\displaystyle \mathop{\delta}_2$ as
$R^\nu_\mu = (R^\nu_\mu)_B + \displaystyle \mathop{\delta}_1 R^\nu_\mu +
\mathop{\delta}_2 R^\nu_\mu$ and $G^\nu_\mu = R^\nu_\mu - {1\over
2}\delta^\nu_\mu R_\gamma^\gamma$. The energy-momentum tensor is
$T^\nu_\mu = \rho u_\mu u^\nu$. In what follows, we assume the 
metric in (\ref{eq:m1}) with conditions (\ref{eq:m8}) and 
(\ref{eq:m9}). For further details see (\cite{acquaviva2003}, 
\cite{maldacena2003}, \cite{nakamura2003}).

For the first order perturbations, we obtain the components of Ricci
and Einstein tensors
\begin{equation}
  \label{eq:a1}
a^2 \mathop{\delta}_1 R^0_0 = 6\Bigl [\Bigl({a'\over a}\Bigr)^2-{a''\over
a}\Bigr ]\psi^{(1)} - \Bigl[\Delta \psi^{(1)}+3 {a'\over a}\psi^{(1)}
+3{\phi^{(1)}}''+3 {a'\over a}\phi^{(1)} \Bigr],
\end{equation}
\begin{equation}
  \label{eq:a2}
a^2 \mathop{\delta}_1 R^i_0 = 2 {\phi^{(1)}_{,i}}' + 2 {a'\over a}
\psi^{(1)}_{,i}, 
\end{equation}
\begin{eqnarray}
  \label{eq:a3}
a^2 \mathop{\delta}_1 G^j_i &=& a^2 [\mathop{\delta}_1 R^j_i -{1\over
2}\delta^j_i \mathop{\delta}_1 R^\mu_\mu] = \Bigl\{2 {a'\over a}{\phi^{(1)}}'
+ \Bigl[{4a''\over a}- 2 \Bigl({a'\over a}\Bigr )^2\Bigr]\psi^{(1)} + \Delta \psi^{(1)} \cr
&+& 2{\phi^{(1)}}'' + {4a'\over a}{\phi^{(1)}}' - \Delta \phi^{(1)}
\Bigr\}\delta_{ij} + \phi^{(1)}_{,ij} - \psi^{(1)}_{,ij}
\end{eqnarray}
and the components of energy-momentum tensor are
\begin{equation}
  \label{eq:a4}
a^2 [\mathop{\delta}_1 T^0_0 - {1\over 2}(\mathop{\delta}_1 T^\mu_\mu)]
= - {1\over 2} \rho_B a^2 \delta^{(1)} \rho/\rho_B,
\end{equation}
\begin{equation}
  \label{eq:a5}
a^2 \mathop{\delta}_1 T^i_0 = - \rho_B a^2 v^{(1)i}
% (a \mathop{\delta}_1 u^i),
\end{equation}
\begin{equation}
  \label{eq:a6}
a^2 \mathop{\delta}_1 T^i_j = 0.
\end{equation}

For the second order perturbations, we obtain the following components 

\begin{eqnarray}
  \label{eq:a7}
a^2 \mathop{\delta}_2 R^0_0 &=& a^2 \mathop{\delta}_2 \tilde{R}^0_0  
 + \psi^{(1)}_{,l}\psi^{(1)}_{,l}
+3{\psi^{(1)}}'{\phi^{(1)}}' -2\phi^{(1)}\Delta\psi^{(1)} \cr
&+& \phi^{(1)}_{,l}\psi^{(1)}_{,l} -6{a'\over a}\phi^{(1)}{\phi^{(1)}}'
-6\phi^{(1)}{\phi^{(1)}}'' 
-3({\phi^{(1)}}')^2  \cr
&+& 2\psi^{(1)} \Bigl[\Delta \psi^{(1)} 
+6{a'\over a}{\psi^{(1)}}' +3{\phi^{(1)}}'' +3{a'\over
a}{\phi^{(1)}}' \Bigr] -12\Bigl[\Bigl({a'\over a}\Bigr)^2 -{a''\over
a}\Bigr] (\psi^{(1)})^2,  
\end{eqnarray}
\begin{equation}
  \label{eq:a8}
a^2 \mathop{\delta}_2 R^i_0 = a^2 \mathop{\delta}_2 \tilde{R}^i_0
-4{a'\over a}\psi^{(1)}\psi^{(1)}_{,i} 
+4{a'\over a}\phi^{(1)}\psi^{(1)}_{,i}
- 2{\phi^{(1)}}' \psi^{(1)}_{,i}+ 4{\phi^{(1)}}'
\phi^{(1)}_{,i} + 8\phi^{(1)} {\phi^{(1)}}'_{,i},  
\end{equation}
\begin{eqnarray}
  \label{eq:a9}
a^2 \mathop{\delta}_2 G^j_i &=& a^2 \mathop{\delta}_2 \tilde{G}^i_j +
W \delta_{ij} + \psi^{(1)}_{,i} \psi^{(1)}_{,j} +2(\psi^{(1)}-
\phi^{(1)}) \psi^{(1)}_{,ij} \cr
&-&\phi^{(1)}_{,i} \psi^{(1)}_{,j}
-\phi^{(1)}_{,j} \psi^{(1)}_{,i} +3\phi^{(1)}_{,i} \phi^{(1)}_{,j}
+4\phi^{(1)} \phi^{(1)}_{,ij}, 
\end{eqnarray}
where
\begin{eqnarray}
  \label{eq:a9w}
W &\equiv& 4\Bigl[\Bigl({a'\over a}\Bigr)^2 -2{a''\over a}\Bigr]
(\psi^{(1)})^2 - 8{a'\over a}\Bigl[\psi^{(1)}{\psi^{(1)}}' +
\psi^{(1)}{\phi^{(1)}}' -\phi^{(1)}{\phi^{(1)}}'\Bigr]
+\Bigl[\phi^{(1)}-2\psi^{(1)}\Bigr]' {\phi^{(1)}}' \cr
&-& \psi^{(1)}_{,l}\psi^{(1)}_{,l} -2\phi^{(1)}_{,l}\phi^{(1)}_{,l}
-4\phi^{(1)} \Delta \phi^{(1)} + 2\Bigl[\phi^{(1)}-\psi^{(1)}\Bigr]
\Bigl[2{\phi^{(1)}}'' + \Delta \psi^{(1)}\Bigr],  
\end{eqnarray}
\begin{equation}
  \label{eq:a10}
a^2 \mathop{\delta}_2 \tilde{R}^0_0 = 3\Bigl[\Bigl({a'\over
a}\Bigr)^2-{a''\over 
a}\Bigr]{\psi^{(2)}} -{1\over 2}\Bigl[\Delta \psi^{(2)}+3 {a'\over a}{\psi^{(2)}}'
+3{\phi^{(2)}}''+3 {a'\over a}{\phi^{(2)}}' \Bigr],
\end{equation}
\begin{equation}
  \label{eq:a11}
a^2 \mathop{\delta}_2 \tilde{R}^i_0 = {\phi^{(2)}}'_{,i} + {a'\over a}
\psi^{(2)}_{,i} + \Bigl[2\Bigl({a'\over a}\Bigr)^2-{a''\over 
a} -{1\over 4}\Delta\Bigr] z_i^{(2)},  
\end{equation}
\begin{eqnarray}
  \label{eq:a12}
a^2 \mathop{\delta}_2 \tilde{G}^j_i &=& 
\Big\{{1\over 2}\Delta \psi^{(2)} +{a'\over a}{\psi^{(2)}}' +
\Bigl[{2a''\over a}- \Bigl({a'\over a}\Bigr)^2\Bigr]\psi^{(2)} +
{\phi^{(2)}}''\cr &+& {2a'\over a}{\phi^{(2)}}' -{1\over 2}
\Delta\phi^{(2)} \Big\}\delta_{ij}  
+ {1\over 2} [\phi^{(2)}_{,ij} - \psi^{(2)}_{,ij}] \cr
&-&{a'\over 2a}[z^{(2)}_{i,j} + z^{(2)}_{j,i}] -{1\over 4}[z^{(2)}_{i,j}
+z^{(2)}_{j,i}]' +{1\over 4}\Bigl[{\chi''_{ij}}^{(2)} + {2a'\over
a}{\chi'_{ij}}^{(2)} -\Delta {\chi_{ij}}^{(2)}\Bigr].
\end{eqnarray}

The latter three components $a^2 \displaystyle 
\mathop{\delta}_2 \tilde{R}^0_0, \
a^2 \mathop{\delta}_2 \tilde{R}^i_0,$ and $a^2 \displaystyle 
\mathop{\delta}_2
\tilde{G}^j_i$ are those which do not
include any second order terms consisting of products of first order 
quantities in $a^2 \displaystyle \mathop{\delta}_1 {R}^0_0, \
a^2 \mathop{\delta}_1 {R}^i_0,$ and $a^2 \displaystyle \mathop{\delta}_1
{G}^j_i$. For energy-momentum tensor, we have
\begin{equation}
  \label{eq:a13}
a^2 \Bigl[\mathop{\delta}_2 T^0_0 - {1\over 2}\mathop{\delta}_2
T^\mu_\mu\Bigr] 
= \rho_B a^2 \Bigl[- {1\over 4} \delta^{(2)} \rho/\rho_B + \sum_i
 (v^{(1)i})^2\Bigr] ,
\end{equation}
\begin{equation}
  \label{eq:a14}
a^2 \mathop{\delta}_2 T^i_0 = - {1\over 2}\rho_B a^2 v^{(2)i}
- a^2(\delta^{(1)} \rho +2\psi^{(1)}\rho_B) v^{(1)i}
- a^2\rho_B \ v^{(1)0} v^{(1)i},
\end{equation}
\begin{equation}
  \label{eq:a15}
a^2 \mathop{\delta}_2 T^i_j = a^2 \rho_B \ v^{(1)i} v^{(1)j}.
\end{equation}
%

% appB
%----------------------------------------------
%\appendix
\section{Second order solutions of Einstein equations}  

In this Appendix we show the solutions of second order Einstein
equations. First, using Eq.(\ref{eq:a9}) in Appendix A and the
expressions of first order perturbations, we obtain 
\begin{eqnarray}
  \label{eq:b1}
\Bigl\{{1\over 2}\Delta \psi^{(2)} &+& {a'\over a}
{\psi^{(2)}}' + \Bigl[{2a''\over a}- \Bigl({a'\over
a}\Bigr)^2\Bigr]\psi^{(2)}  
+ {\phi^{(2)}}'' + {2a'\over a}{\phi^{(2)}}' -{1\over 2}
\Delta\phi^{(2)} \Bigr\} \delta_{ij}  
+ {1\over 2} [\phi^{(2)}_{,ij} - \psi^{(2)}_{,ij}] \cr
&-& {a'\over 2a}[z^{(2)}_{i,j}
+ z^{(2)}_{j,i}] -{1\over 4}[z^{(2)}_{i,j}
+z^{(2)}_{j,i}]' +{1\over 4}\Bigl[\chi''_{ij} + {2a'\over a}\chi'_{ij}
-\Delta \chi_{ij}\Bigr] + W\delta_{ij} \cr
&=& \Bigl\{\Bigl[\Bigl({a' \over a}\Bigr)^2 -{a'' \over
2a}\Bigr](P')^2 -{1 \over 2}\Bigl(1 - {a' \over a}P'\Bigr)^2\Bigr\} 
F_{,i} F_{,j} - \Bigl(1 - {a' \over a}P'\Bigr)^2 F F_{,ij}.
\end{eqnarray}
Here $F F_{,ij} = {1 \over 2}(F^2)_{,ij} - F_{,i}F_{,j}$ and
$F_{,i}F_{,j}$ is divided into following independent terms
\begin{equation}
  \label{eq:b2}
F_{,i}F_{,j} = A\delta_{ij} + B_{,ij} +C_{i,j} +C_{j,i} +D_{ij},
\end{equation}
where the following conditions are imposed on $C_i$ and $D_{ij}$
\begin{equation}
  \label{eq:b3}
\delta^{ij}C_{i,j} = 0, \ \delta^{ij} D_{ij} = 0 \ {\rm and} \
\delta^{jk} D_{ij,k} = 0.
\end{equation}
Functions $A, B, C_i$ and $D_{ij}$ are determined as follows
\begin{equation}
  \label{eq:b4}
A = {100 \over 9} \Psi_0,
\end{equation}
\begin{equation}
  \label{eq:b5}
B = -{100 \over 3} \Theta_0,
\end{equation}
\begin{equation}
  \label{eq:b6}
\Delta C_i = \Bigl[{200 \over 9} \Psi_0 - {1\over
2}F_{,l}F_{,l}\Bigr]_{,i} + F_{,i} \Delta F,
\end{equation}
\begin{equation}
  \label{eq:b7}
D_{ij} = F_{,i}F_{,j} -{100 \over 3}\Psi_0 \delta_{ij} + {100 \over 3}
\Theta_{0,ij} - (C_{i,j}+C_{j,i}).
\end{equation}
Corresponding to the relation (\ref{eq:b2}), Eq.(\ref{eq:b1}) is
divided into four parts
\begin{eqnarray}
  \label{eq:b8}
2{a''\over a}\psi^{(2)} &+& {a'\over a} {\psi^{(2)}}' + {1\over 2}\Delta
\psi^{(2)}- \Bigl({a'\over a}\Bigr)^2 \psi^{(2)} +{\phi^{(2)}}''+  {2a'\over a}
{\phi^{(2)}}' - {1\over 2}\Delta \phi^{(2)} \cr
&=& -\Bigl\{\Bigl[\Bigl({a'\over a}\Bigr)^2 -{2a''\over a}\Bigr]
\Bigl(1-{a'\over a}P'\Bigr)^2 +{2a'\over 
a} \Bigl(1-{a'\over a}P'\Bigr)\Bigl({a'\over a}P'\Bigr)' -{1\over
4}\Bigl[\Bigl({a'\over a}P'\Bigr)'\Bigr]^2\Bigr\} 
F^2 \cr
&+& {1\over 4} \Bigl(1-{a'\over a}P'\Bigr)^2 [2\Delta (F^2) - F_{,l}F_{,l}] +
\Bigl\{\Bigl[\Bigl({a'\over a}\Bigr)^2 -{a''\over 2a}\Bigr] (P')^2 +
{1\over 2} \Bigl(1-{a'\over a}P'\Bigr)^2\Bigl\} {100\over 9} \Psi_0,
\end{eqnarray}
\begin{equation}
  \label{eq:b9}
\psi^{(2)} -\phi^{(2)} = \Bigl(1-{a'\over a}P'\Bigr)^2 F^2 +
\Bigl\{1-{2 a'\over a}P' 
+\Bigl[3\Bigl({a'\over a}\Bigr)^2 -{a''\over a}\Bigr](P')^2\Bigr\}
{100\over 3} \Theta_0, 
\end{equation}
\begin{equation}
  \label{eq:b10}
{z_i^{(2)}}' + {2a'\over a}z_i^{(2)} = -2 \Bigl\{1-{2a'\over a}P'
+\Bigl[3\Bigl({a'\over a}\Bigr)^2 -{a''\over a}\Bigr](P')^2 \Bigr\}  C_i, 
\end{equation}
\begin{equation}
  \label{eq:b11}
{\chi^{(2)}_{ij}}'' + {2a'\over a}{\chi^{(2)}_{ij}}' - \Delta
\chi^{(2)}_{ij} = {1\over 2}\Bigl\{1-{2a'\over a}P'
+\Bigl[3\Bigl({a'\over a}\Bigr)^2 
-{a''\over a}\Bigr](P')^2 \Bigr\} D_{ij}. 
\end{equation}
Eliminating $\psi^{(2)}$ from Eqs.(\ref{eq:b8}) and (\ref{eq:b9}), we
obtain an equation for $\phi^{(2)}$
\begin{eqnarray}
  \label{eq:b12}
{\phi^{(2)}}'' &+&{3a'\over a}{\phi^{(2)}}' + \Bigl[{2a''\over
a}-\Bigl({a'\over a}\Bigr)^2\Bigr] \phi^{(2)} \cr
&=& {1\over 4}\Bigl[\Bigl({a'\over a}\Bigr)' P' +{a'\over
a}\Bigl(1-{2a'\over a}P'\Bigr)\Bigr]^2 F^2 
+{1\over 4}\Bigl\{1 -{2a'\over a} P' +\Bigl[5\Bigl({a'\over a}\Bigr)^2
-{2a''\over a}\Bigr] (P')^2 \Bigr\} F_{,l}F_{,l} \cr
&+& \Bigl\{3\Bigl({a'\over a}\Bigr)^2 -{2a''\over a} +{a'\over
a}\Bigl[{8a''\over a}- 14\Bigl({a'\over a}\Bigr)^2\Bigr] P' +
\Bigl[ 2\Bigl({a''\over a}\Bigr)^2 + 19 \Bigl({a'\over a}\Bigr)^4\cr
&-& 14 \Bigl({a'\over a}\Bigr)^2 {a''\over a} \Bigr]
(P')^2 \Bigr\} {100 \over 3}\Theta_0 
- \Bigl\{1 - {2a'\over a} P' + \Bigl[3\Bigl({a'\over a}\Bigr)^2 
- {a''\over a}\Bigr] (P')^2 \Bigr\} {100\over 9} \Psi_0.  
\end{eqnarray}
Solving this equation and using Eq.(\ref{eq:b9}), we get the
expressions for $\phi^{(2)}$ and $\psi^{(2)}$ given in
Eqs.(\ref{eq:m16}) - (\ref{eq:m17}). 

% appC
%----------------------------------------------
%\appendix
\section{The expression of $\Psi_0$ and the integration of
$(F_{,r})^2$, $\Psi_0$, and $F$ along a light path}

The term $I$ in Eq.(\ref{eq:c19}) is
\begin{eqnarray}
  \label{eq:p1}
I &=& {1\over 18}c^2r^2  \qquad \quad {\rm for} \quad 0 \le r \le r_0 \cr
&=& {1\over 12}r_0^2(c-3b)(c+b) -{1\over 36}r^2 [(c+b)^2(r_0/r)^6
-8(c+b)b(r_0/r)^3 -2b^2]  \qquad {\rm for} \quad r_0 < r \le r_1 \cr
&=& {100\over 9}\Psi_0(0) -{1\over 36}[c r_0^3 -b(r_1^3-r_0^3)]^2 r^{-4} \qquad
{\rm for} \quad r_1 < r, 
\end{eqnarray}
where 
\begin{eqnarray}
  \label{eq:p2}
{100\over 9}\Psi_0(0) &\equiv& {1\over 12}r_0^2(c-3b)(c+b) -{1\over
36}r_1^2 [(c+b)^2(r_0/r_1)^6 -8(c+b)b(r_0/r_1)^3 -2b^2] \cr
&-& {1\over 36}[cr_0^3 -b(r_1^3-r_0^3)]^2 r_1^{-4}. 
\end{eqnarray}
For $0 < r_* <r_0$, we obtain the following integrals by use of $u
\equiv r/r_0$ and $y \equiv b/c$ 
\begin{eqnarray}
  \label{eq:p3}
\int^\infty_0 (F_{,r})^2 dx &/&({1\over 18}c^2r_0^3) = -4y(1+y) \ln {u_1 +
(u_1^2 -u_*^2)^{1/2} \over 1 + (1- u_*^2)^{1/2}} +{2\over
3}(1-u_*^2)^{1/2} (1-y^2)(1+2u_*^2) \cr
&+& {2\over 3}(u_1^2-u_*^2)^{1/2} y^2(u_1^2 +2u_*^2) 
+(1+y)^2 u_*^{-2} \Bigl\{(u_1^2-u_*^2)^{1/2}/u_1^2 - (1-u_*^2)^{1/2} \cr
&+& u_*^{-1} \tan^{-1} {u_*[(u_1^2 -u_*^2)^{1/2}- (1- u_*^2)^{1/2}]
\over u_*^2 + [(u_1^2 -u_*^2)(1- u_*^2)]^{1/2}} \Bigr\} + 
[1+y(1-u_1^3)]^2 u_*^{-2} [-(u_1^2-u_*^2)^{1/2}/u_1^2 \cr
&+& {1\over u_*} \tan^{-1} {u_* \over (u_1^2 -u_*^2)^{1/2}}]
\end{eqnarray}
and
\begin{eqnarray}
  \label{eq:p4}
{100\over 9} \int^\infty_0 \Psi_0 dx &/&({1\over 18}c^2r_0^3) =
-4y(1+y) \ln {u_1 + (u_1^2 -u_*^2)^{1/2} \over 1 + (1- u_*^2)^{1/2}}
-{3\over 2}(1-u_*^2)^{1/2} \Bigl[{2\over 9}(1-y^2)(1+2u_*^2) \cr
 &+& (3y-1)(1+y)\Bigr] - {3\over 2}(u_1^2-u_*^2)^{1/2} 
\Bigl[{2\over 9}y^2(u_1^2
+2u_*^2)  
-2y(1+y)/u_1 -y^2 u_1^2\Bigr ] \cr
&+& {1\over 4}(1+y)^2 u_*^{-2}
\Bigl\{(u_1^2-u_*^2)^{1/2}/u_1^2 - (1-u_*^2)^{1/2}  \cr
&+& u_*^{-1} \tan^{-1} {u_*[(u_1^2 -u_*^2)^{1/2}- (1- u_*^2)^{1/2}]
\over u_*^2 + [(u_1^2 -u_*^2)(1- u_*^2)]^{1/2}} \Bigr\} + {1\over 4}
[1+y(1-u_1^3)]^2 u_*^{-2} \Bigl[-(u_1^2-u_*^2)^{1/2}/u_1^2 \cr
&+& {1\over u_*}\tan^{-1} {u_* \over (u_1^2 -u_*^2)^{1/2}}\Bigr],
\end{eqnarray}
where $u_1 \equiv r_1/r_0, u_* \equiv r_*/r_0$ and $x \equiv (r^2
-r_*^2)^{1/2}$. 
For $r_* = 0$, we obtain the compensating case (i.e.,
${u_1}^3 = 1+ 1/y$)
\begin{eqnarray}
  \label{eq:p5}
\int^\infty_0 (F_{,r})^2 dr &=& {100\over 9} \int^\infty_0 \Psi_0 dr \cr
&=& {2\over 27} c^2 r_0^3 \ (1+y)(1 - 3y \ln u_1).
\end{eqnarray}

The integration of $F$ is derived as follows :
\begin{equation}
  \label{eq:p6}
\int^\infty_0 F dx = \int^{r_1}_{r_*} F \Bigl(r^2 - r_*^2\Bigr)^{-1/2}
r dr 
\equiv c(r_0)^3 J\Bigl(u_* \Bigr), 
\end{equation}
where
\begin{eqnarray}
  \label{eq:p7}
J\Bigl(u_* \Bigr)
&=& -{1\over 9} \Bigl(1-{u_*}^2 \Bigr)^{1/2}(1+y)\Bigl(4-{u_*}^2\Bigr)
+{1\over 9} \Bigl({u_1}^2 -{u_*}^2 \Bigr)^{1/2} \Bigl[y \Bigl({u_1}^2
-{u_*}^2 \Bigr) +3(1+y)/u_1 \Bigr] \cr
&-& {1\over 3} (1+y) \ln {u_1 +\Bigl({u_1}^2 - {u_*}^2\Bigr)^{1/2}
\over 1 +\Bigl(1 - {u_*}^2\Bigr)^{1/2}}.
\end{eqnarray}
In the above derivation we assumed that the mass is totally
compensating. For $r_* = 0$, we obtain 
\begin{equation}
  \label{eq:p8}
J(0) = -{1\over 3} (1+y) \ln \Bigl(u_1 \Bigr) \  
= -{1\over 9} (1+y) \ln (1 +1/y).
\end{equation}
%  

%ref
%\begin{references}

%*****************************************************

\end{document}